\documentclass[lettersize,journal]{IEEEtran}
\usepackage{amsmath,amsfonts}
\usepackage{algorithmic}
\usepackage{algorithm}
\usepackage{array}
\usepackage{textcomp}
\usepackage{stfloats}
\usepackage{url}
\usepackage{verbatim}
\usepackage{graphicx}
\usepackage[utf8]{inputenc}
\usepackage[style=numeric,sorting=none]{biblatex}
\usepackage{subcaption}
\begin{document}

\title{ACM with Overlapping Partitions: Implementation and Periodicity Analysis}

\author{Anthony O'Dea\\Department of Chemical Engineering,\\University of California, Santa Barbara \vspace{-2em} % <-this % stops a space
\thanks{Acknowledgment}% <-this % stops a space
\thanks{Manuscript submission and review/acceptance date}}



\maketitle

\begin{abstract}
The Arnold Cat Map (ACM) is a popular chaotic map used in image encryption. Chaotic maps are known for their sensitivity to initial conditions and their ability to mix, or rearrange, pixels. However, ACM is periodic, and the period is relatively short. This periodicity decreases the effective key space for a cryptosystem. Further, ACM can only be performed on square matrices. For non-square images, this issue can be solved by performing ACM on multiple square partitions of the image. If these partitions overlap, the periodicity will greatly increase. The resulting system will be referred to as overlapping ACM or OACM. This paper will cover the implementation and periodicity analysis for these overlapping systems, which previous papers involving similar overlapping block partitions did not. Viewing OACM as a scan as opposed to a map allows for faster implementation and period analysis.
\end{abstract}


\section{Introduction to ACM}
The Arnold cat map is an area preserving linear transformation. ACM is one of several chaotic maps used in image encryption. ACM is mainly used to rearrange an image's pixels such that after some number of operations, the resulting image will appear randomized. ACM can also be used to alter an image's RGB values. ACM in matrix form is: 
\begin{equation}
\begin{bmatrix} x' \\ y' \end{bmatrix} = 
\begin{bmatrix} 1 & 1 \\ 1 & 2 \end{bmatrix} 
\begin{bmatrix} x \\ y \end{bmatrix} 
\bmod N = 
\begin{bmatrix} A \end{bmatrix} 
\begin{bmatrix} x \\ y \end{bmatrix} 
\bmod N
\end{equation}
[A] is shorthand for the ACM matrix. For image encryption, N will typically equal the side length in pixels of a square image. For decryption, the inverse of the ACM matrix is used to move pixels back to their original state. Equation 1 may be rewritten as: %
\begin{equation} x' = (x + y) \bmod N ,\, y' = (x + 2y) \bmod N \end{equation} 
In this form, it is non-obvious why one can't just use ``modulo width'' and ``modulo height'' to find x' or y' to extend ACM to all rectangular images. The reason is that for general rectangular images, the modulo operation in eq. (2) can map multiple pairs of x and y to the same x' and y'. However, it is possible to use ``modulo width'' and ``modulo height'' without issue when the aspect ratio is S:1, with S as a positive integer. This is often ignored in literature, and will be ignored for the rest of this paper. 

ACM is generalized quite frequently as:
\begin{equation}
\begin{bmatrix} x' \\ y' \end{bmatrix} = 
\begin{bmatrix} 1 & p \\ q & 1 + pq \end{bmatrix} 
\begin{bmatrix} x \\ y \end{bmatrix} 
\bmod N
\end{equation}
Where p and q are integer values ranging from 0 to N - 1. In this paper, ACM will refer to the original ACM operation, where p and q are both equal to 1. Other generalizations and modifications of ACM exist, but they must have a determinant of plus or minus unity to remain area preserving \cite{tribonacci,generalizedACM}. 
\IEEEpubidadjcol

ACM is performed some number of iterations, Z, on an image during the encryption process. The value Z may be indicated within the key. Rather than performing ACM Z times on an image, it would be ideal to have a single operation which would provide the same output image. Multiple iterations of Arnold cat map may simply be represented via:
\begin{equation}
\begin{bmatrix} x^Z \\ y^Z \end{bmatrix} = 
\begin{bmatrix} A \end{bmatrix}^Z 
\begin{bmatrix} x \\ y \end{bmatrix} 
\bmod N
\end{equation}
However, the values in ${[A]^{Z}}$ quickly become large enough to cause numerical errors, unless otherwise accounted for. To avoid these large values, an equivalent matrix, ${A_{Z}}$, can be found with the following iterative sequence:
\begin{equation}
\begin{bmatrix} A_{n} \end{bmatrix} = 
\begin{bmatrix} A_{n-1} \end{bmatrix} 
\begin{bmatrix} A \end{bmatrix} 
\bmod N
\end{equation}
This sequence is periodic, and the period, P(N), is a function of N. Certain relations have been found between N and P(N):
\begin{align*} 
P(N) &= 3N \, for \, N = 2*5^k \\
P(N) &= 2N \, for \, N =5^k \, or \, 6*5^k \\
P(N) &<= \tfrac {12}{7} N \, for \, other \, N
\end{align*}%
With k as positive integers. The periodicity never exceeds 3N \cite{periodRels}.  For common image sizes, the periodicity is considerably smaller than this maximum (table 1). 
\begin{table}[!h]
\begin{center}
\caption{Regular ACM data}
\label{tab1}
\begin{tabular}{ c | c  c  c  c  c  c  c  c}
Size & 256 & 512 & 1024 & 1080 & 2048 & 2992 & 3000 & 4320  \\
Period & 192 & 384 & 768 & 180 & 1536 & 180 & 1500 & 360  \\
\end{tabular}
\end{center}
\end{table}

Periodicity limits the effective keyspace and increases susceptibility of the cryptosystem to brute force attacks. Cryptosystems involving chaotic maps with weak keyspaces have been broken before \cite{breaking}. Several modifications to ACM have already been proposed in literature to increase periodicity. The simplest of which involve modifying p and q. ACM can also be extended to 3D \cite{symmetric}. Nonlinear perturbations can also be added to ACM \cite{linear}. 

A pixel within an image undergoing ACM has its own period, which is the minimum number of iterations before returning to its original location. The path a pixel takes during these iterations is referred to as its orbit. All of the pixels along an orbit have the same period, which is equal to the number of pixels in that orbit. The (0, 0) pixel does not move, so it has a period of 1. The overall P(N) of an image will be the least common multiple of the individual pixel periods. 

At certain iterations of ACM before P(N), the original image may reappear in some manner \cite{ghosts}. These iterations will be referred to as recurrence periods and typically are simple fractions of P(N) (figure 1). Recurrence periods may be of some concern regarding image security, but are not usually considered as a limitation to possible keyspaces within other literature.

\begin{figure}[!h]
\centering
\includegraphics[width=1.6in]{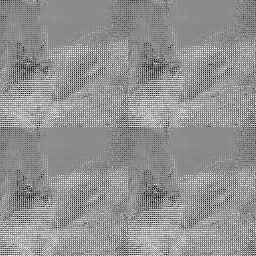}
\, \,
\includegraphics[width=1.6in]{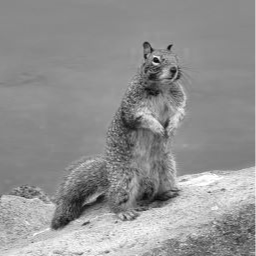}
\caption{256 X 256 Squirrel image after P(256)/2 iterations vs original.}
\label{fig_1}
\end{figure}

\section{Overlapping ACM}

Previous works have proposed the used of overlapping block partitions, but they focused on parameters regarding the block partitions rather than the impact on periodicity \cite{projPart,spiral,acmOverlap,blockShuffle}. Overlapping block partitions can extend ACM to non-square images and will greatly increase the overall image period. Without restrictions, there are an infinite number of ways to lay squares on a rectangle. This may be desirable in a cryptosystem, but is not conducive to studying how overlapping partitions influence periodicity. Therefore, this paper utilizes a method which produces overlapping partitions similar to a regular grid. 

There are several key parameters to this problem: image height and width, square size, amount of overlap, and number of squares. The number of squares is determined by the other parameters and can not be directly set within the proposed method detailed below. 

The method begins by producing a list of x and y coordinates for the upper left corner of each square, based on the given parameters. The respective tiling will be made with the least number of squares possible. The specified amount of overlap is sometimes broken at the edges of the image, in cases where the tiling would not cover the entire image. The coordinates are found via algorithm 1. 

\begin{algorithm}[!h]
\caption{Square Locations.}\label{alg:alg1}
\begin{algorithmic}
\STATE 
\STATE {{for y from 0 to image height - square size - 1 by 1's}}
\STATE \hspace{0.5cm}if (y-1) Mod (square size - overlap) = 0\\
\STATE \hspace{0.5cm}\hspace{0.5cm}add y to y list \\
\STATE \hspace{0.5cm} end
\STATE  end
\STATE add height-squareSize to y list
\STATE
\STATE {{for x from 0 to image width - square size - 1 by 1's}}
\STATE \hspace{0.5cm}if (x-1) Mod (square size - overlap) = 0\\
\STATE \hspace{0.5cm}\hspace{0.5cm}add x to x list \\
\STATE \hspace{0.5cm} end
\STATE  end
\STATE add width-squareSize to x list
\end{algorithmic}
\label{alg1}
\end{algorithm}

%this call prevents the second column from running into the IEEE statement % at the bottom of the page
ACM is then performed on each square in order of top left to bottom right. Decryption is performed via inverse ACM on each square in the reverse order.

A limiting/simplifying case is when a rectangle is to be tiled by the minimum number of squares of size equalling the image height. For rectangles with aspect ratio W:H between 1:1 and 2:1, this tiling is simply one square on the leftmost and rightmost sides. This limiting case will be explored to further introduce the proposed method. For squares, a simplifying case would be a tiling with four squares in each corner with edge lengths \(>\) half the image size. 

Tables II and III indicate the overall period for simplified rectangular and square OACM systems respectively. The periods within these tables were found as the least common multiple of the orbit lengths.

\begin{table}[!h]
\begin{center}
\caption{Period data for rectangles tiled with two squares of size H}
\label{tab2}
\begin{tabular}{ c | c | c | c }
%\hline
W:H & W & H & Period \\
\hline

16:9 & 1920 & 1080 &  $9.2*10^{489}$ \\
%\hline
3:2 & 1620 & 1080 & $1.7*10^{77}$ \\
%\hline
32:27 & 1280 & 1080 & $1.8*10^{330}$ \\
%\hline
3:2  & 1536  & 1024 & $1.6*10^{104}$ \\
%\hline
5:4 & 1280 & 1024 & $1.7*10^{369}$ \\
%\hline
4:3 & 1280 & 960 & $1.3*10^{332}$ \\
%\hline
16:9 & 1600 & 900 & $4.9*10^{387}$ \\
%\hline
4:3 & 1024 & 768 & $2.9*10^{251}$ \\
%\hline
16:9 & 1280 & 720 & $6.6*10^{344}$ \\
%\hline
3:2 & 1080 & 720 & $4.0*10^{37}$ \\
%\hline
4:3 & 960 & 720 & $2.2*10^{171}$ \\
%\hline
5:4 & 900 & 720 & $5.3*10^{219}$ \\
%\hline
4:3 & 768 & 576 & $2.8*10^{147}$ \\
5:4 & 720 & 576 & $8.4*10^{98}$ \\
%\hline
4:3 & 640 & 480 & $1.7*10^{167}$ \\
%\hline
4:3 & 320 & 240 & $1.9*10^{81}$ \\
%\hline
\end{tabular}
\end{center}
\end{table}

\begin{table}[!h]
\begin{center}
\caption{Period data for square images tiled with 4 overlapping squares}
\label{tab3}
\begin{tabular}{ c | c | c }
%\hline
Image Size & Square Size & Period\\
\hline
1080 & 720 & $7.5*10^{165}$ \\
%\hline
1024 & 768 & $1.1*10^{458}$ \\
%\hline
1024 & 640 & $5.9*10^{504}$ \\
%\hline
768 & 512 & $3.7*10^{262}$ \\
512 & 384 & $8.1*10^{236}$ \\
%\hline
256 & 192 & $4.0*10^{106}$ \\
%\hline
%\hline
\end{tabular}
\end{center}
\end{table}

\section{Implementing Overlapping ACM} 
Unlike ACM, multiple iterations of overlapping ACM (OACM) can not be represented with a single matrix multiplication. Performing OACM multiple times on an image instead of as a single operation would be very disadvantageous in terms of computation time. This issue can be solved by pre-computing the pixel orbits. This will essentially treat OACM as a ``scan'' as opposed to a chaotic map. Scans are another popular image encryption technique which involve rearranging pixels based on alternate methods of scanning through an image. 

The orbits of every pixel can be derived from just a single operation of OACM. This operation is used to generate a matrix where the value at every pixel is the index of the next location along its orbit. Moving to that next location yields another point along the orbit, and so on. Once all the orbits are known, the location of a pixel after Z iterations of OACM will be moved along the orbit by ``Z modulo orbit length'' positions. This method should work with other chaotic maps or scans.

\section{Periodicity Analysis}
The large increases in image periodicity are a direct result of increases in the length of orbits and the number of unique orbit lengths (figure 2). Note that increasing the ``complexity'' of a system by increasing the number of squares will not necessarily result in a higher image periodicity. Further, certain combinations of square size and overlap can result in very long orbit lengths. This limits the effective period to the length of that orbit. This effective period will be referred to as a recurrence period, even though it is somewhat different from regular ACM's recurrence periods. 

\begin{figure}[!h]
\centering
\includegraphics[width=3.25in]{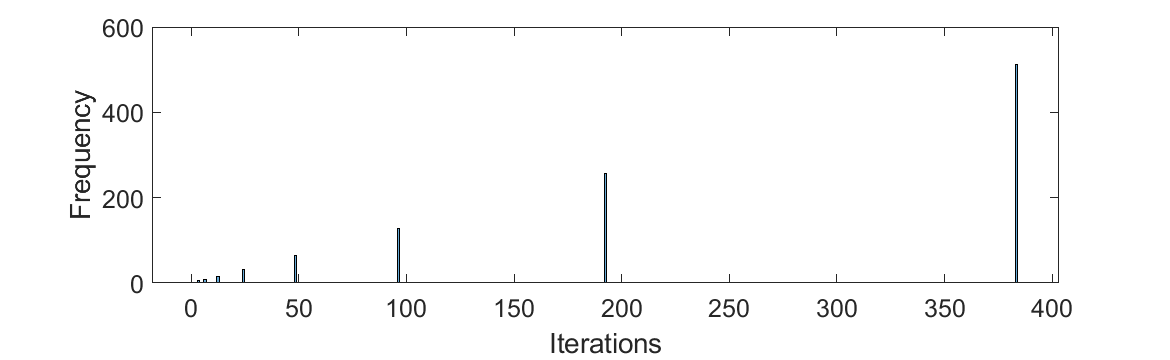}
\includegraphics[width=3.25in]{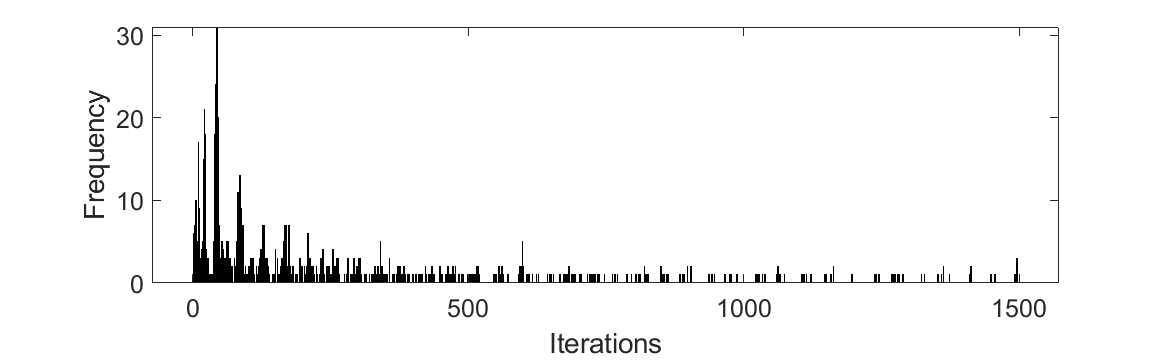}
\includegraphics[width=3.25in]{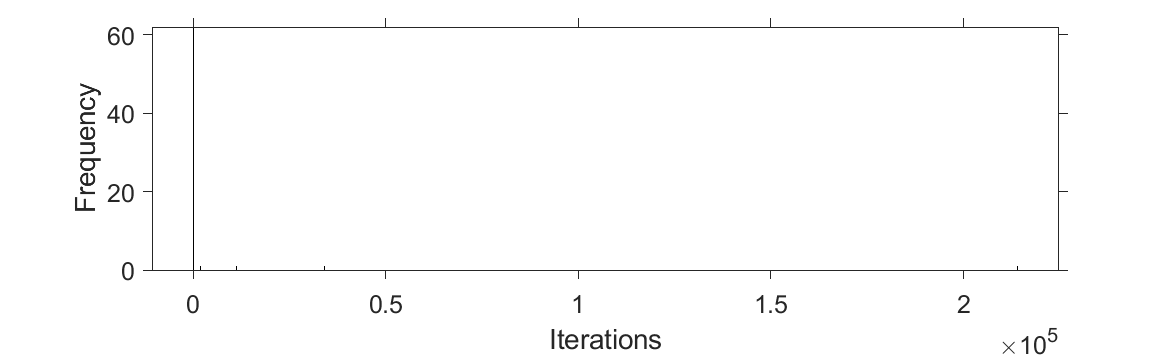}
\caption{Histograms indicating the frequency of orbit lengths. From top to bottom, regular ACM, OACM with 4 squares of size 384, OACM with squares of size 100 and an overlap of 25. Note that shorter orbits tend to occur more frequently than large orbits in OACM, but their total number of pixels is still negligible compared to the larger orbits.}
\label{fig_2}
\end{figure}

Recurrence periods may be found graphically using similarity graphs (figure 3). Similarity graphs represent the number or percentage of pixels back in their original location vs the number of iterations. Noticeable peaks in these graphs will occur when multiple orbits return to their original state at the same time or when an orbit of sufficient length returns to its original state.

\begin{figure}[!h]
\centering
\includegraphics[width=3.25in]{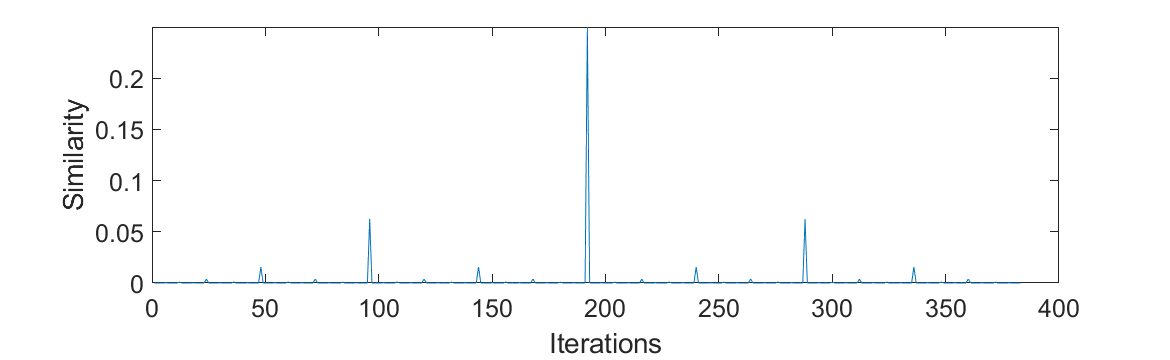}
\includegraphics[width=3.25in]{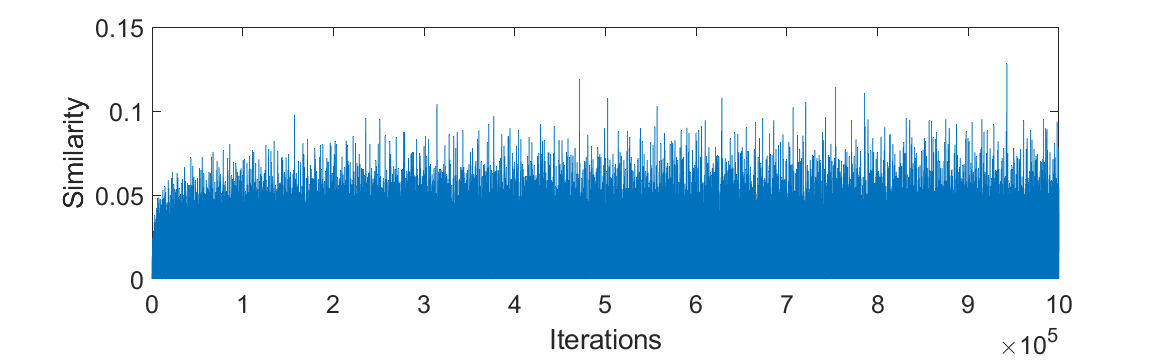}
\includegraphics[width=3.25in]{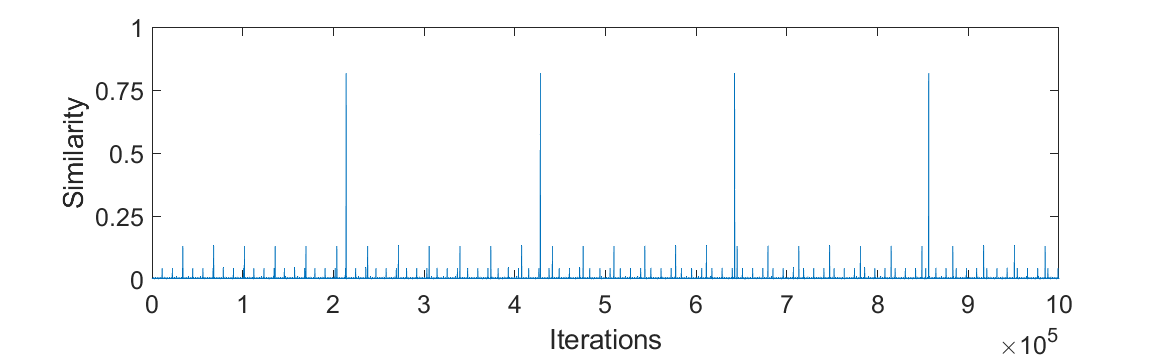}
\caption{Similarity graphs for 512 by 512 images. From top to bottom, regular ACM, OACM with 4 squares of size 384, OACM with squares of size 100 and an overlap of 25. Note the presence of peaks above 75\% similarity in the bottom graph, which are resultant from a single large orbit}
\label{fig_3}
\end{figure}

%\begin{figure}[!h]
%\centering
%\begin{subfigure}{3.25in}
%    \includegraphics[width=3.25in]{512regularACMSimGraph.png}
%    \subcaption{}
%    \label{fig:first}
%\end{subfigure}
%\begin{subfigure}{3.25in}
%    \includegraphics[width=3.25in]%{512with384foursquaresimgraph.png}
%    \subcaption{}
%    \label{fig:second}
%\end{subfigure}
%\begin{subfigure}{3.25in}
 %   \includegraphics[width=3.25in]{512with100and25oversimgraph.png}
 %   \subcaption{}
 %   \label{fig:third}
%\end{subfigure}

%\caption{Similarity graphs for 512 by 512 images. From top to %bottom, regular ACM, OACM with 4 squares of size 384, OACM with %squares of size 100 and an overlap of 25. }
%\label{fig_4}
%\end{figure}

Given that OACM provides some level of control over the orbit lengths, what set of these lengths will yield the maximum possible image period? A similar question was posed by Edmund Landau around the start of the 20th century \cite{landauIntro}: what partitions of n elements will provide the maximal order? The LCM of the lengths of the partitions is the order, and would also be maximized. The ``Landau series,'' s(n), for n elements will be a set of partition lengths which provide maximal order. There may be multiple s(n) for the same n. The Landau function, g(n), provides the order for s(n). Note that the Landau function is not a singular complete function, and there are several existing methods for finding g(n) with varying speeds and ranges of accuracy \cite{landauEstimate,landauOneMillion}. The prime factorization of g(n) may be used to find s(n) where any recurring primes within the prime factor list are replaced by their product. 

The values of g(n), with n as the number of pixels, provide an upper bound for OACM's and similar systems' periodicity. For example, a 512 by 512 image with Landau sized partitions would have a period of g(262144), equalling $4.3*10^{826}$. Note the much smaller OACM period values in tables II and III.

\section{Conclusion}
OACM has much higher periods than regular ACM and allows for the encryption of non-square images. However, these periods are much lower than the theoretical maximum provided by the Landau function. Treating OACM as a scan allows for much faster implementation than when treated as multiple ACM operations. The scan method should work for other chaotic maps as well, whether or not they overlap. 

Future related work may involve extending OACM to 3D, using partitions which wrap around an image's boundaries, and using partitions of aspect ratio S:1 with S as a positive integer. Developing encryption techniques based on the Landau series is also of interest for future work.

%\section*{Acknowledgments}
%I would like to acknowledge Dr. Çetin Kaya Koç, Dr. Eric McFarland, and Dr. James Rawlings for their support and advice during the writing of this paper.

%\section*{Data Availability}
%The main OACM programs are available here:  . 

%\begin{thebibliography}{1}
\printbibliography

\newpage

%% the LaTeX parser from getting confused when it sees the complicated
% $\backslash${\tt{includegraphics}} command within an optional argument. %(You can create
% your own custom macro containing the $\backslash${\tt{includegraphics}} command to make things
 
\vspace{11pt}

%{\appendix[Proof of the Zonklar Equations]
%Use $\backslash${\tt{appendix}} if you have a single appendix:
%Do not use $\backslash${\tt{section}} anymore after %$\backslash${\tt{appendix}}, only $\backslash${\tt{section*}}.
%If you have multiple appendixes use $\backslash${\tt{appendices}} then use %$\backslash${\tt{section}} to start each appendix.
%You must declare a $\backslash${\tt{section}} before using any %$\backslash${\tt{subsection}} or using $\backslash${\tt{label}} %($\backslash${\tt{appendices}} by itself
 %starts a section numbered zero.)}

\vfill

\end{document}